\documentclass[preprint,authoryear,12pt]{elsarticle}

\usepackage{amssymb}
\usepackage{color}

\journal{Earth and Planetary Science Letters}

\begin{document}

\begin{frontmatter}



\title{Evolution of Water Reservoirs on Mars:\\ Constraints from Hydrogen Isotopes in Martian Meteorites}


\author[label1,label2]{H. Kurokawa}\ead{kurokawa@nagoya-u.jp}
\author[label3,label2]{M. Sato}
\author[label2]{M. Ushioda}
\author[label2]{T. Matsuyama}
\author[label2]{R. Moriwaki}
\author[label4]{J. M. Dohm}
\author[label2]{T. Usui}

\address[label1]{Department of Physics, Nagoya University, Furo-cho, Chikusa-ku, Nagoya, Aichi 464-8602, Japan}
\address[label2]{Department of Earth and Planetary Sciences, Tokyo Institute of Technology, 2-12-1 Ookayama, Meguro, Tokyo 152-8551, Japan}
\address[label3]{Department of Environmental Changes, Kyushu University, 744 Motooka, Nishi-ku, Fukuoka 819-0395, Japan}
\address[label4]{Earth-Life-Science Institute, Tokyo Institute of Technology, 2-12-1-1E-1 Ookayama, Meguro-ku, Tokyo, 152-8550, Japan}

\begin{abstract}
Martian surface morphology implies that Mars was once warm enough to maintain persistent liquid water on its surface.
While the high D/H ratios ($\sim 6$ times the Earth's ocean water) of the current martian atmosphere suggest that significant water has been lost from the surface during martian history, 
the timing, processes, and the amount of the water loss have been poorly constrained. 
Recent technical developments of ion-microprobe analysis of martian meteorites have provided accurate estimation of hydrogen isotope compositions (D/H) of martian water reservoirs at the time when the meteorites formed.
Based on the D/H data from the meteorites, this study demonstrates that the water loss during the pre-Noachian ($> 41-99\ {\rm m}$ global equivalent layers, GEL) was more significant than in the rest of martian history ($> 10-53\ {\rm m\ GEL}$).
Combining our results with geological and geomorphological evidence for ancient oceans, 
we propose that undetected subsurface water/ice ($\simeq 100-1000\ {\rm m\ GEL}$) should have existed, and it exceeds the observable present water inventory ($\simeq 20-30\ {\rm m\ GEL}$) on Mars.
\end{abstract}

\begin{keyword}
Mars \sep meteorites \sep water reservoir \sep isotope \sep atmospheric escape
\end{keyword}

\end{frontmatter}


\section{Introduction} \label{}

Mars is generally considered to be a cold and dry planet,
with relatively small amounts of water-ice observed at the polar caps \citep[e.g.,][]{Jakosky+Phillips2001,Christensen+2006}. 
On the contrary, 
a number of geological observations, such as dense valley networks \citep{Scott+1995,Carr+Chuang1997,Hoke+2011} and deltas \citep{Cabrol+Grin1999,Ori+2000,DiAchille+Hynek2010}, provide definitive evidence that large standing bodies of liquid water (i.e., oceans and lakes) existed in the early history, 
the presence of which would have profound implications for the early climate and habitability of Mars \citep[e.g.,][]{Carr2007,Head+1999,Dohm+2011}. 
The geological observations further include the detection of water-lain sediments and a variety of hydrous minerals (e.g., clays) \citep[e.g.,][]{Fialips+2005,Bibring+2006} and evaporites (e.g. gypsum) \citep[e.g.,][]{Osterloo+2008} commonly formed by aqueous processes, 
implying Earth-like hydrologic activities, with Noachian lakes and/or oceans. 
Despite such compelling evidence for hydrologic conditions that could support oceans and lakes, 
there are, however, major gaps in our understanding of the evolution of surface water: 
e.g., what was the global inventory of martian surficial water/ice, and how did it change through time.

The global inventory of ancient surficial water has been estimated based on the size of reported paleo-oceans \citep[e.g.,][]{Head+1999,Clifford+Parker2001,Carr+Head2003,Ormo+2004,DiAchille+Hynek2010}. 
Topographic features of putative paleo-shorelines suggest that large bodies of standing water once occupied the northern lowlands \citep{Head+1999}. 
Shoreline-demarcation studies of the northern lowlands point to several contacts that yield variable sizes of paleo-oceans estimated to range from $\sim 2 \times 10^7\ {\rm km^3}$ to $2 \times 10^8\ {\rm km^3}$ (corresponding to global equivalent layers (GEL) of $130\ {\rm m}$ to $1,500\ {\rm m}$, respectively) \citep[][and references therein]{Carr+Head2003}. 
Though the shoreline demarcations \citep{Parker+1989,Parker+1993}, 
supported by Mars Orbiter Laser Altimeter (MOLA) topography \citep{Head+1998,Head+1999}, 
could not be confirmed using Mars Orbiter Camera (MOC) and Thermal Emission Imaging System (THEMIS) image data \citep{Malin+Edgett1999,Malin+Edgett2001,Ghatan+Zimbelman2006}, 
this variation has been interpreted to reflect the historical change in the ocean volume. 
For example, two major contacts (contact-1: Arabia shoreline and contact-2: Deuteronilus shoreline) individually represent the larger Noachian and smaller Hesperian oceans, respectively \citep{Parker+1993,Clifford+Parker2001,Carr+Head2003}. 
Although these geomorphologic studies have provided significant constraints on the history of martian paleo-oceans, 
they lack information about pre-Noachian \citep{Frey2006} oceans because no geologic records are available. 
Furthermore, the shoreline-demarcation approaches would not be applicable to the youngest Amazonian (3.1 Ga to present) era, during which the surface water would have occurred mostly as ice \citep{Clifford+Parker2001,Carr+Head2010}. 

This study endeavors to trace the global inventory of surficial water through time beginning with the embryonic stages of development of Mars (i.e., 4.5 Ga) to present day based on a geochemical approach of hydrogen isotopes (D/H: deuterium/hydrogen). 
Hydrogen is a major component of water (${\rm H_2O}$) and its isotopes fractionate significantly during hydrological cycling between the atmosphere, surface water, and ground and polar cap ices. 
Telescopic studies have reported that the hemispheric mean of the martian atmosphere has a D/H ratio of ~6 times (${\rm \delta D} \simeq 5000$ñ) the terrestrial values \citep{Owen+1988}; ${\rm\delta D = [(D/H)_{sample}/(D/H)_{reference} -1] \times 1000}$, where the reference is Standard Mean Ocean Water (SMOW). 
Because the high atmospheric D/H ratio is interpreted to result from the preferential loss of hydrogen relative to the heavier deuterium from the martian atmosphere throughout the planetfs history \citep{Lammer+2008}, 
the deuterium enrichment can be used to estimate the amount of water loss due to the atmospheric escape. 

Compared to a number of geomorphologic studies \citep[e.g.,][]{Scott+1995,Head+1999,Clifford+Parker2001,Carr+Head2003,DiAchille+Hynek2010}, 
only a few geochemical investigations have been conducted \citep{Chassefiere+Leblanc2011,Lammer+2003}. 
This is partly because there have been a limited number of reliable D/H datasets for martian meteorites, and the martian meteorites typically have younger ages (typically, $< 1.3\ {\rm Ga}$) \citep{Nyquist+2001}. 
However, recent technical developments of ion-microprobe analysis of martian meteorites including the 4.1 Ga ALH 84001 pyroxenite have provided more accurate and comprehensive datasets for D/H ratios of martian water reservoirs \citep[e.g.,][]{Greenwood+2008,Usui+2012,Hallis+2012a,Hallis+2012b}, 
yielding new information helpful for unraveling the origin and evolution of water on Mars. 
Furthermore, 
although martian meteorites were derived from limited and highly biased regions of the surface of Mars \citep{McSween+2009,Usui+2010,Christen+2005}, 
their radiometric ages are more accurate and precise than crater counting ages.

Based on the recent D/H dataset from martian meteorites, 
we estimate the amount of water loss during $4.5\ {\rm Ga}$ to $4.1\ {\rm Ga}$, which we refer to here as Stage-1 in our analysis (approximating the pre-Noachian; Table 1),
and consequently demonstrate that the water loss during $4.5\ {\rm Ga}$ to $4.1\ {\rm Ga}$ was more significant than in the rest of the Mars history ($4.1\ {\rm Ga}$ to present, approximating Noachian - Amazonian; Table 1), 
which we refer to as Stage-2 in our analyses.
Combining our results with geological estimates for the volume of martian paleo-oceans, 
we propose that unidentified surficial water-ice reservoirs should currently exist and the volume ($\simeq 100 - 1000\ {\rm m\ GEL}$) should exceed the estimated present water inventory \citep[20-30 m GEL,][]{Christensen+2006} on Mars.

\section{Calculation}

The amount of water loss due to the atmospheric escape between time $t_1$ and $t_2$ can be calculated from an assumed amount of present water reservoir using the following equations:
\begin{equation}
L_{t_1 - t_2} = R_{t_1} - R_{t_2} = R_{t_2} \times \Biggl[ \Biggl( \frac{I_{t_2}}{I_{t_1}} \Biggr)^{\frac{1}{1-f}} - 1 \Biggr] , \label{EstimateLoss}
\end{equation}
and
\begin{equation}
f = {\rm \frac{d [D]/[D]}{d [H]/[H]}}.
\end{equation}
Here $L_{t_1 - t_2}$ is the amount of water loss due to the atmospheric escape during the time from $t_1$ to $t_2$, 
$R$ and $I$ are an amount of water reservoir and a D/H ratio at each time, respectively, $f$ is the fractionation factor, 
and ${\rm [H]}$ and ${\rm [D]}$ are the abundances of H and D in the combined reservoirs in ${\rm atoms\ cm^{-2}}$  \citep{Lammer+2003}.
Both the volumes of water reservoir and water loss are expressed in ocean depth [m] as a global equivalent layer (GEL).
Using the density of water of $10^3\ {\rm kg\ m^{-3}}$ and the surface area of $1.4\times 10^{14}\ {\rm m^2}$, 
$1\ {\rm m}$ GEL corresponds to $1.4\times 10^{17}\ {\rm kg}$ of water.
Eq. \ref{EstimateLoss} can be rewritten as:
\begin{equation}
\frac{R_{t_2}}{R_{t_1}} = \Biggl( \frac{I_{t_2}}{I_{t_1}} \Biggr)^{\frac{1}{1-f}},\label{Rrelation}
\end{equation}
which gives a ratio of the amount of water for $t_1$ and $t_2$.
Eq. \ref{Rrelation} is used in Sections 3.2, 3.3, and 3.4 to discuss the evolution of the amount of water through time.

We employ the fractionation factor $f$ of $0.016$, a representative value reported for the present martian condition \citep{Krasnopolsky+1998,Krasnopolsky2000}. 
Since the D/H fractionation due to magmatic outgassing is relatively insignificant \citep[e.g., $f = 0.9$,][]{Pineau+1998}, 
we consider the D/H fractionation to be solely due to atmospheric escape.
We calculate surface water loss in two stages: Stage-1 ($4.5-4.1\ {\rm Ga}$) and -2 ($4.1\ {\rm Ga}$ - present), as shown in Fig. 1. 
The boundary ($4.1\ {\rm Ga}$) is derived from the crystallization age of ALH 84001, the only martian meteorite which records a Noachian crystallization age \citep{Lapen+2010}, 
though there has been a very recent reporting of a martian meteorite NWA 7533 with an extremely ancient ($\simeq 4.4\ {\rm Ga}$), pre-Noachian age \citep{Humayun+2013}. 

The D/H ratios of the surficial water reservoir $I$ at the boundary conditions ($4.5\ {\rm Ga}$, $4.1\ {\rm Ga}$, and the present) are obtained from hydrogen isotope analyses of martian meteorites. 
The initial ${\rm \delta D}$ value of $275$ñ for the $4.5\ {\rm Ga}$ primordial martian water is obtained from analyses of olivine-hosted melt inclusions from a primitive basaltic meteorite, Yamato 980459 (shergottite). 
This meteorite represents a primary melt from a depleted mantle source formed at $\sim 4.5\ {\rm Ga}$, and its melt inclusions are interpreted to possess undegassed primordial water in the martian mantle \citep{Usui+2012}. 
The ${\rm \delta D}$ range ($1200-3000$ñ) of the near-surficial water reservoir at $4.1\ {\rm Ga}$ is derived from analyses of magmatic phosphate and secondary carbonate minerals in ALH 84001 \citep{Boctor+2003,Greenwood+2008}.
The ${\rm \delta D}$ value ($5000$ñ) of the present martian water reservoir is obtained from D/H analyses of geochemically enriched shergottites (Shergotty and LAR 06319) that crystallized near the surface in the recent past \citep[$0.17-0.18\ {\rm Ga}$,][]{Greenwood+2008,Usui+2012}.
This high ${\rm \delta D}$ value is consistent with those of water in the present martian atmosphere ($\simeq 5000$ñ) determined through both telescopic \citep[e.g., ][]{Owen+1988} and the {\it Curiosity} rover observations \citep{Webster+2013}.

D/H ratios of martian meteorites reflect complex geologic histories including the terrestrial weathering after the meteorite falls. 
This study employs D/H datasets obtained only from recent {\it in situ} ion microprobe measurements to minimize the effect of terrestrial contamination. 
The ALH 84001 carbonates formed by ancient aqueous activity possess surficial water components, 
whereas the D/H ratios of magmatic phosphates (apatites) are interpreted as representing mixing of \textquotedblleft magmatic" and \textquotedblleft surficial" water components \citep{Boctor+2003}. 
However, as D/H ratios of apatites in geochemically enriched shergottites are indistinguishable from that of the current martian atmosphere, 
they are interpreted as representing a D/H ratio of near-surface water that is exchanged with the atmosphere (previously called the exchangeable reservoir) \citep{Greenwood+2008}. 
The D/H ratios of apatites employed in this study approximate the surficial water D/H ratios when the host meteorites crystallized near the surface. 

\section{Results and Discussions}

\subsection{Estimate of water loss}

Using Eq. \ref{EstimateLoss}, 
the estimated water loss in Stage-1 ($L_{\rm 4.5-4.1 Ga}$) and Stage-2 ($L_{\rm 4.1-0 Ga}$) is obtained as a function of the amount of the present water reservoir $R_{\rm present}$ (Fig. 2).
The ranges of $L_{\rm 4.5-4.1 Ga}$ and $L_{\rm 4.1-0 Ga}$ at a given $R_{\rm present}$ (i.e., width of the stripes in Fig. 2) reflect the range of the D/H ratios at $4.1\ {\rm Ga}$ ($1200-3000$ñ).
The results show that $L_{\rm 4.5-4.1 Ga}$ and $L_{\rm 4.1-0 Ga}$ are positively correlated with $R_{\rm present}$.

Our model further indicates that a ratio of water loss between the stages ($L_{\rm 4.5-4.1 Ga} / L_{\rm 4.1-0 Ga}$) is independent of $R_{\rm present}$ and that $L_{\rm 4.5-4.1 Ga}$ is always greater than $L_{\rm 4.1-0 Ga}$ at any $R_{\rm present}$.
Dividing Eq. \ref{EstimateLoss} for Stage-2 by that for Stage-1, the ratio can be written as,
\begin{equation}
\frac{L_{\rm 4.5-4.1 Ga}}{L_{\rm 4.1-0 Ga}} = \frac{\Bigl(\frac{I_{\rm 0Ga}}{I_{\rm 4.1Ga}}\Bigr)^{\frac{1}{1-f_2}}\Bigl[\Bigl(\frac{I_{\rm 4.1Ga}}{I_{\rm 4.5Ga}}\Bigr)^{\frac{1}{1-f_1}}-1\Bigr]}{\Bigl(\frac{I_{\rm 0Ga}}{I_{\rm 4.1Ga}}\Bigr)^{\frac{1}{1-f_2}}-1},
\end{equation}
where $f_1$ and $f_2$ are the fractionation factors in both Stage-1 and -2.
Assuming the same fractionation factor $f$ for both Stage-1 and -2, the ratio of water loss is given by
\begin{equation}
\frac{L_{\rm 4.5-4.1 Ga}}{L_{\rm 4.1-0 Ga}} = \frac{I_{4.5 {\rm Ga}}^{-\frac{1}{1-f}} - I_{4.1 {\rm Ga}}^{-\frac{1}{1-f}}}{I_{4.1 {\rm Ga}}^{-\frac{1}{1-f}} - I_{0 {\rm Ga}}^{-\frac{1}{1-f}}}. \label{ratioL}
\end{equation}
This equation shows that the ratio of water loss is determined only from the D/H ratios and the fractionation factor in Stage-1 and -2.
Because the ${\rm \delta D}$ value of $1200-3000$ñ at $4.1\ {\rm Ga}$ is already fractionated from the initial D/H ratio of $275$ñ, 
the ratio $L_{\rm 4.5-4.1 Ga}/L_{\rm 4.1-0 Ga}$ given by Eq. \ref{ratioL} is $\simeq 1.2 - 6.5$.
This indicates that the water loss is more significant in Stage-1 when compared to Stage-2. 
As the period of Stage-1 ($0.4\ {\rm Gyrs}$) is $\simeq 10$ times shorter than Stage-2 ($4.1\ {\rm Gyrs}$), 
the average escape rate in Stage-1 would be more than $10$ times higher than in Stage-2.

\subsection{Minimum estimate of water loss}

The amount of the \textquotedblleft observable" current surface water reservoir is dominated by the polar layered deposits (PLD).
Assuming that the PLDs are mainly composed of water-ice, 
they are expected to contain ${\rm H_2O}$ of $1.2 - 1.6\ \times 10^6\ {\rm km^3}$ in the North polar region \citep{Zuber+1998} and $1.6\ \times 10^6\ {\rm km^3}$ in the South polar region \citep{Plaut+2007}, respectively; 
their total sum ($2.8 - 3.2 \times 10^6\ {\rm km^3}$) corresponds to $20-30\ {\rm m}$ GEL.
We employ this value ($20-30\ {\rm m}$) as the minimum estimate for the amount of present water reservoir $R_{\rm present}$, 
because the existence of \textquotedblleft missing" water-ice reservoirs has been proposed \citep[e.g.,][]{Carr+Head2003}.
For example, ice-rich mantles and covering sediments in the mid-latitude possibly contain a large amount of ice \citep{Murray+2005,Page2007,Christensen+2006,Page+2009}.
Furthermore, there is increasing evidence that vast reservoirs of water-ice potentially exist in parts of the high latitudes, 
as indicated by geomorphology \citep{Baker2001,Kargel2004,Soare+2007,Soare+2011,Soare+2012,Soare+2013a,Soare+2013b,Smith+2009,Lefort+2009,Levy+2009a,Levy+2009b,Levy+2011}, {\it in situ} analysis through the Phoenix Lander \citep{Smith+2009}, and the Mars Odyssey Gamma Ray Spectrometer \citep{Boynton+2002,Boynton+2007}.

As there is a positive correlation between the amount of present water reservoir and the water loss (Fig. 2), 
the minimum $R_{\rm present}$ of $20 - 30\ {\rm m}$ GEL yields the minimum estimate of the water loss in each stage (Fig. 2). The minimum water loss in Stage-1 ($L_{\rm 4.5-4.1Ga}$) and -2 ($L_{\rm 4.1-0Ga}$) are calculated to be $41 - 99\ {\rm m}$ GEL and $10 - 53\ {\rm m}$ GEL, respectively (Table 2). 
The sum of these values ($R_{\rm present}$, $L_{\rm 4.1-0Ga}$, $L_{\rm 4.5-4.1Ga}$) yields the minimum estimates for the amounts of martian water reservoirs of $30-83\ {\rm m}$ GEL at $4.1\ {\rm Ga}$ and $97-150\ {\rm m}$ at $4.5\ {\rm Ga}$, respectively (Fig. 3). 
The ranges for these estimates are derived from the uncertainty of the amount of present water reservoir ($20 - 30\ {\rm m}$) and the uncertainty of the D/H ratio at $4.1\ {\rm Ga}$ ($1200-3000$ñ).
The required average escape rates of water molecule to explain the calculated water loss in both stages are $1.6 - 3.8\ \times 10^{28}\ {\rm molecules\ s^{-1}}$ in Stage-1 and $0.38 - 2.0\ \times 10^{27}\ {\rm molecules\ s^{-1}}$ in Stage-2, respectively.

We employ the fractionation factor $f$ of $0.016$, which is valid under present martian conditions \citep{Krasnopolsky+1998,Krasnopolsky2000}.
The D/H fractionation is mainly induced by two different mechanisms.
One is the escape-induced;  fractionation in the uppermost atmosphere where hydrogen tends to escape relative to heavier deuterium.
The other is the fractionation induced by photochemical reactions in the lower atmosphere; hydrogen molecules are depleted in deuterium compared to water molecules as a result of a series of photochemical reactions to convert water molecules to hydrogen molecules \citep[e.g.,][]{Yung+1988}.
High extreme UV (EUV) radiation of the younger Sun, which induces high exobase temperature \citep{Kulikov+2007}, 
reduces the former D/H fractionation by the atmospheric escape as a result of intense escape of both H and D. 
Thus, because the $f$ of $0.016$ employed in this study is likely to be minimum (i.e., largest fractionation), 
our model yields the minimum estimate of global water loss. 
Even if it is granted, our main conclusion, more water loss in Stage-1 than Stage-2, would not change, 
because $f$ is thought to be greater in the older Stage-1 than in the younger Stage-2 due to the hotter exobase condition in Stage-1 than in Stage-2.

Our model does not take into account the effects of water supply by magmatic outgassing and by late accretion of water-bearing bodies such as asteroids and comets. 
Prolonged igneous activities would have delivered magmatic water from the martian interior to the surface. 
Because such magmatic water is assumed to have unfractionated primordial hydrogen isotopic compositions \citep[e.g., $\simeq 275$ñ,][]{Usui+2012}, 
the magmatic outgassing is expected to suppress the hydrogen isotope fractionation of the surficial and atmospheric water reservoirs. 
Thus, more water should have been lost than our estimates to explain the D/H ratios at the boundary conditions of $4.5\ {\rm Ga}$ and $4.1\ {\rm Ga}$; i.e., our model again provides the minimum estimate for the amounts of the surface-water loss.

Hydrogen isotopic compositions of asteroids and comets are variable. 
The extreme cases are carbonaceous-chondrite parent asteroids \citep[e.g., from $-230$ to $+340$ñ,][]{Alexander+2012} and Oort-cloud comets \citep[$\sim 1000$ñ,][]{Hartogh+2011}, respectively. 
Because the delivery of low-D/H water by late accretion of such chondritic bodies tends to maintain lower hydrogen isotope compositions of the surficial water reservoirs, 
its effect on our calculation should be similar to the effect of magmatic outgassing discussed above. 
On the other hand, cometary impacts can increase the D/H ratio of the surface water reservoir without atmospheric escape. 
For example, the supply of $10^{19}\ {\rm kg}$ comets (corresponding to $\sim 100\ {\rm m}$ GEL) with a D/H ratio of $1000$ñ increases the D/H ratio of surface water reservoir by $\sim 1000$ñ. 
However, comets are typically enriched in noble gases along with water-ice. 
Late accretion of $10^{19}\ {\rm kg}$ comets with a probable ${\rm Xe/H_2O}$ ratio of $\sim 10^{-5}$ \citep{Swindle2012} results in a supply of $10^{14}\ {\rm kg}$ ${\rm Xe}$, 
which is $10^6$ times larger than the martian atmospheric ${\rm Xe}$. 
Since it is almost impossible for such a large amount of ${\rm Xe}$ to escape during $4\ {\rm Gyrs}$, 
such a significant supply of comets is unlikely.

\subsection{Comparison with geological records}

Our model requires an amount of surface water at a specific age to calculate amounts of surface water at other ages for which the meteorite data are available. 
In the previous section, 
we have reported the minimum volumes for the surface waters at $4.5\ {\rm Ga}$ and $4.1\ {\rm Ga}$, respectively, 
by assuming the present water inventory ($R_{\rm present}$) of $20-30\ {\rm m}$ GEL. 
However, as noted earlier, PLD represents the minimum estimate for the present water inventory and the existence of \textquotedblleft undetected" current water-ice reservoirs has been proposed. 
Thus, through utilizing geological estimates for the volume of ancient oceans, 
this section examines the transition of volumes of Mars ocean from $4.5\ {\rm Ga}$ to the present.

Water bodies ranging in size from lakes to oceans possibly occupied the martian landscape, corroborated through growing evidence, which includes deltas at certain topographic zonal boundaries \citep[e.g.,][]{DiAchille+Hynek2010}, 
giant polygons similar to those found on the ocean floors of Earth \citep{McGowan2011}, 
and impact craters and associated features reminiscent of impacts into Earth oceans \citep{Ormo+2004}. 
In addition, the existence of ancient oceans is supported by Mars Odyssey Gamma Ray Spectrometer-based elemental information, 
which highlights distinctions between the regions below and above the putative shorelines \citep{Dohm+2009}, 
long-wavelength topography consistent with deformation caused by true polar wandering \citep{Perron+2007}, 
and Mars Advanced Radar for Subsurface and Ionosphere Sounding (MARSIS) data which indicates that regionally occurring fine-grained sediments such as marine deposits and/or ice rather than lava flows best explain the radar signatures \citep{Mouginot+2012}.

Geologic and geomorphologic studies have proposed a wide range of estimates for the volumes of Mars paleo-oceans \citep[e.g.,][]{Head+1999,Clifford+Parker2001,Fairen+2003,Carr+Head2003,Ormo+2004,DiAchille+Hynek2010}. 
Shoreline demarcation studies of the northern plains have resulted in the identification of two contrasting ocean-forming inundations \citep{Parker+1987,Parker+1993}, 
and their estimated spatial extents \citep{Head+1999,Carr+Head2003}. 
Contact-1, redefined of greater extent as the Arabia shoreline, 
records the larger ocean which may have occurred at a certain period during the Early Noachian until possibly through the Early Hesperian \citep{Fairen+2003}, 
or $\simeq 4.0\ {\rm Ga}$ to $3.6\ {\rm Ga}$ based on the cratering model of \citet{Hartmann+Neukum2001}. 
On the other hand, Contact-2, identified as Deuteronilus shoreline, 
represents the smaller ocean at a certain period during Late Hesperian through the Early Amazonian \citep{Fairen+2003}, or $\simeq 3.6\ {\rm Ga}$ to $1.4\ {\rm Ga}$ based on the model of \citet{Hartmann+Neukum2001}. 
\citet{Carr+Head2003} proposed a younger and smaller ocean with a volume of 160 m GEL in the Late Hesperian ($3 - 3.5\ {\rm Ga}$), 
by examining the spatial distribution of Vastitas Borealis Formation (VBF) and the crater-retention age of VBF sediments. Based on the global distribution of deltas and valleys, \citet{DiAchille+Hynek2010} proposed a Noachian ocean with a volume of 550 m GEL and argued that it would have an age of $\simeq 3.5\ {\rm Ga}$ (Noachian/Hesperian boundary).

We compile these geological estimates for the volumes of ancient oceans and compare them with our model calculations (Fig. 4).
Our estimates reproduce a general trend of the geological records that Noachian oceans have greater amounts of water than Hesperian oceans.
However, our minimum estimate based on the PLD does not quantitatively match these geological estimates.
The discrepancy of our minimum estimate with the geological records suggests the existence of unidentified present water reservoirs ($\simeq 100 - 1000\ {\rm m\ GEL}$), which exceed the total PLDs observed ($20 - 30\ {\rm m\ GEL}$) (Fig. 4). 
Such missing water-ice reservoirs might be mid-latitude ice mantles and ice-rich sediments whose amounts are not well constrained, or under-ground ice implied by {\it Mars Express}'s dielectric mapping observations \citep{Mouginot+2012}.
Note that an inefficient fractionation (larger $f$) can also explain the great water loss without a large D/H fractionation, though it requires an unrealistic atmospheric condition or other escape mechanisms such as impact erosion.

The comparison also suggests that the initial water amount at $4.5\ {\rm Ga}$ would be a range of $\sim 10^2 - 10^3\ {\rm m}$ GEL, 
which corresponds to $\sim 10^{-2} - 10^{-1}\ {\rm wt}\ \%$ of total Mars mass.
Formation theories of terrestrial planets suggest a wide range of the water mass fraction of bulk Mars because the initial condition is unknown and the formation process involves stochastic properties.
\citet{Lunine+2003} calculated the accretion of Mars, assuming that planetary embryos initially range in a wide position of orbital radii.
They obtained the water mass fraction of Mars to be $0.014 - 0.063\ {\rm wt\ \%}$.
More recently, in the framework of the \textquotedblleft Grand Tack" scenario \citep{Walsh+2011} in which planetary embryos range in a more compact position, 
\citet{Brasser2013} estimated the water mass fraction of Mars to be $0.1 - 0.2\ {\rm wt\ \%}$.
Though the surficial water discussed in this study is a part of the total martian water inventory and the amount of water in the mantle is unknown, 
our estimate of the initial water amount ($\sim 10^{-2} - 10^{-1}\ {\rm wt\ \%}$ of the Mars mass) is consistent with these theoretical predictions.

\subsection{Constraints for oxygen sinks}

The \textquotedblleft bottleneckh to restrict the water loss is remaining oxygen as a result of the hydrogen escape. 
As oxygen escapes mainly due to the interaction of the uppermost atmosphere with solar wind, 
the amount of oxygen escape is influenced by the presence and intensity of martian magnetic field. 
Fig. 3 and Table 2 compare our minimum estimates for water loss in Stages-1 and -2 with that estimated based on oxygen escape models. 
An oxygen escape model by \citet{Lammer+2003} proposes the water loss of $24-58\ {\rm m}$ GEL during the recent $3.5\ {\rm Gyr}$ under the assumption that Mars did not possess a global magnetic field; this assumption is consistent with the absence of crustal magnetism in post-Noachian geologic units \citep{Dohm+2013}. 
Intense magnetic anomalies observed over the Noachian highlands indicate the presence of a martian dynamo before $\simeq 4\ {\rm Ga}$ \citep{Lillis+2008}.
However, it has been disputed whether the dynamo had been active during the entire period of Stage-1 because of the lack of geologic records before 4.2 Ga \citep{Frey2006}. 
If the magnetic protection was weak or absent, strong winds from the young sun would have induced greater escape than that of today. \citet{Terada+2009} calculated the wind-induced oxygen escape in the earliest $150\ {\rm Myr}$ of the martian history, which corresponds to a possible saturation phase of the solar wind, 
and obtained water loss of $31 - 133\ {\rm m}$ GEL without the magnetic protection in the pre-Noachian period. 
Our minimum estimates of water escape ($L_{\rm 4.5-4.1 Ga} = 41-99\ {\rm m}$ GEL and $L_{\rm 4.5-4.0Ga} = 10-53\ {\rm m}$ GEL) are consistent with these oxygen escape models. 
This consistency indicates that no- or weak- dynamo would have contributed to the significant water loss during pre-Noachian.

Calculated water loss based on any geological estimates of paleo-oceans is distinctly greater than our estimate based on the PLD. 
In one instance, this study employs the Late Hesperian VBF ocean \citep{Carr+Head2003} to calculate the amount of water loss. 
Because it should have existed in a period within Stage-2 ($4.1$ Ga to the present), 
we assume that the VBF ocean had a $\delta {\rm D}$ value between $1200$ñ (the lowest end at $4.1$ Ga) and $5000$ñ (the present value). 
Using Eq. \ref{Rrelation} to estimate the amounts of water reservoirs both forward and backward in time, 
the evolution of the water reservoir is estimated (Table 2). 
The estimated water loss based on the VBF ocean is much larger than the water loss estimated from the oxygen escape calculations \citep{Lammer+2003,Terada+2009}. 
As these oxygen escape calculations are considered a maximum estimate without the magnetic protection induced by dynamo field, 
the discrepancy indicates that there might have been other mechanisms of oxygen loss. 

Oxidation of surface materials is another possible mechanism to account for the remaining excess oxygen.
\citet{McSween+2006} reported fresh basalt compositions for Adirondack-class basalts on the Gusev plains.
Using these compositions, we estimate how much rock can be oxidized by the excess oxygen.
Assuming that ${\rm Fe^{2+}}$ and ${\rm S^{2-}}$ in the fresh basalts are oxidized to ${\rm Fe^{3+}}$ and ${\rm S^{6+}}$ 
and that the density ratio of rock to water as $3:1$, 
$14.8\ {\rm m\ GEL}$ of the fresh basalts can consume oxygen in $1\ {\rm m\ GEL}$ of water. 
If ${\rm CaO}$ and ${\rm MgO}$ are also oxidized to form ${\rm CaSO_4}$ and ${\rm MgSO_4}$, 
$0.99\ {\rm m}$ of the fresh basalts is required to consume oxygen in $1\ {\rm m}$ GEL of water. 
Such mass balance calculations suggest that the surface oxidation process appears insufficient to account for the greater water loss (up to $520\ {\rm m}$ GEL, Table 2) estimated based on the VBF ocean, 
because at least the equivalent volume of basaltic crust (i.e. $\simeq 500\ {\rm m}$ GEL) should be required. 
This implies the existence of unknown oxygen escape mechanisms or unrevealed oxygen sinks. 

\section{Conclusion}

Geological and geomorphological studies have revealed that Mars once contained large amounts of liquid water on its surface. 
We estimate the amount of water loss due to atmospheric escape in two stages based on the D/H data of martian meteorites. 
We demonstrate that the amount of water loss is positively correlated with the present water inventory and that the water loss during $4.5\ {\rm Ga}$ to $4.1\ {\rm Ga}$ (Stage-1) is more significant than that in the rest of the martian history (Stage-2), regardless of the amount of the present water reservoir. 
Adopting the minimum estimate of the present water inventory based on the estimated extent of the PLD yields the minimum estimates of water loss. 
The minimum estimates of water loss are comparable with those obtained from the oxygen escape calculations.
Combining our results with geological constraints for ancient oceans, we propose a possibility that there should be undetected subsurface water/ice of much greater extent than the collective amounts of the \textquotedblleft visibleh current water inventory. 
Our study further implies that, because such a large water inventory automatically calls for significant water loss that cannot be explained either by the existing oxygen escape models or the known oxygen sinks, unknown mechanisms that effectively consume the remaining excess oxygen are required. 

\section*{Acknowledgments}

We acknowledge T. Ikoma and H. Genda for fruitful discussions.
This work was supported by the program for the \textquotedblleft Global Center of Excellence for the 21st Century in Japan" to Department of Earth and Planetary Sciences, Tokyo Institute of technology, 
and by \textquotedblleft 2013 Tokyo Institute of Technology Challenge Research Award" and by a NASA Mars Fundamental Research Program grant to TU.
HK is supported by Grants-in-Aid from the Ministry of Education, Culture, Sports, Science and Technology (MEXT) of Japan (23244027).

\clearpage

\begin{table}
\begin{center}
\includegraphics[scale=1.0]{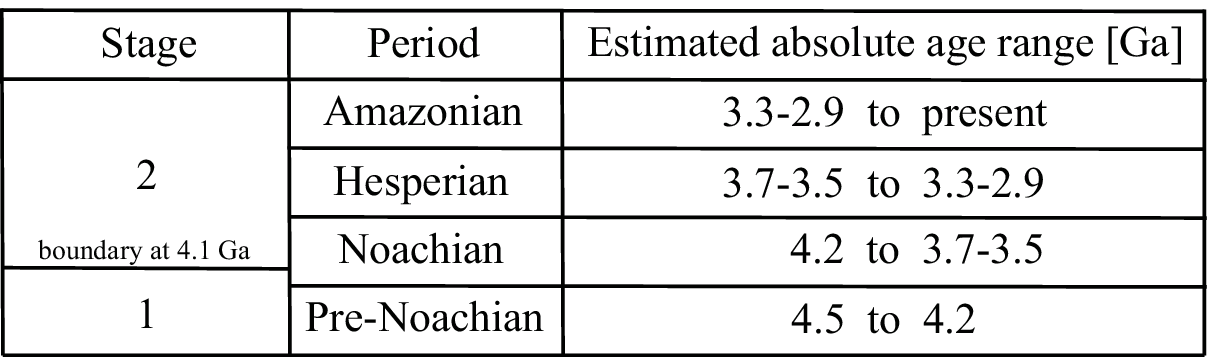}
\caption{
Geologic-mapping-based, time-stratigraphic information, including the Noachian, Hesperian, and Amazonian Periods \citep{Scott+Carr1978}.
The geologic periods have been given estimated absolute age ranges based on impact crater models (Table 1 is modified from \citet{Hartmann+Neukum2001}). 
Note that we performed comparative analyses among the estimated water amounts of the pre-Noachian, 
approximately referred to here as Stage-1, and Noachian-Amazonian, approximately referred to as Stage-2, 
based on recent D/H dataset from martian meteorites.
}
\end{center}
\end{table}

\clearpage

\begin{figure}
\begin{center}
\includegraphics[scale=1.0]{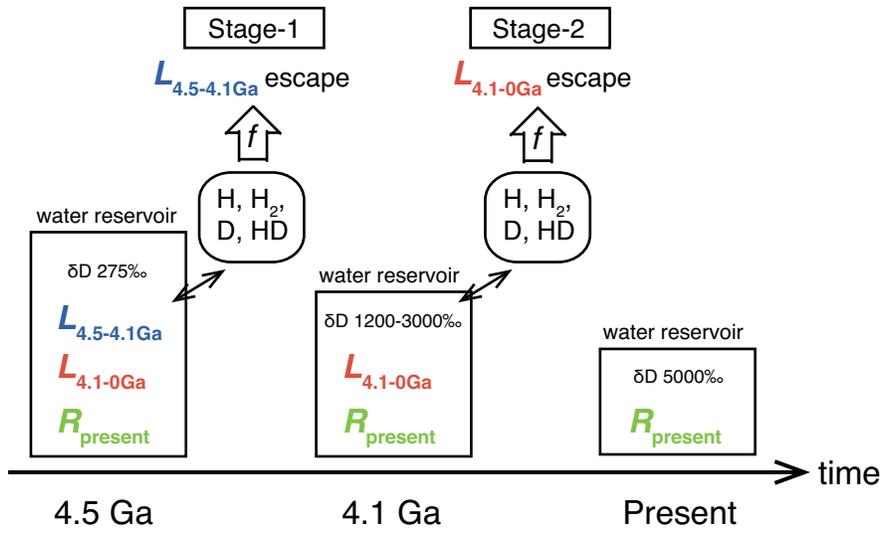}
\caption{
Schematic illustration of the two-stage model for the evolution of the global surface water reservoir on Mars. 
$R_{\rm present}$ is the size of the present water reservoir, 
$L_{\rm 4.5-4.1 Ga}$ and $L_{\rm 4.1-0 Ga}$ are the water loss during Stage-1 and -2, 
and $f$ is the fractionation factor (see text).
}
\end{center}
\end{figure}

\clearpage

\begin{figure}
\begin{center}
\includegraphics[scale=1.0]{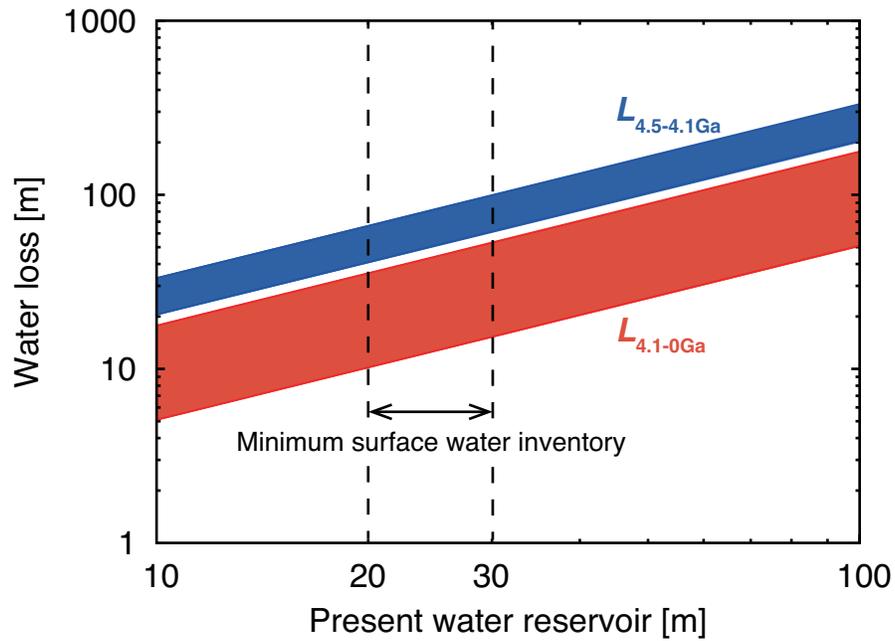}
\caption{
Water loss during Stage-1 ($L_{\rm 4.5-4.1Ga}$, blue) and Stage-2 ($L_{\rm 4.1-0Ga}$, red) as a function of present water reservoir $R_{\rm present}$. 
The width of the blue and red stripes is derived from the ${\rm \delta D}$ range ($1200 - 3000$ñ) for ALH 84001. 
The range of the minimum estimate for the present water reservoir in PLD (see text in detail) is bracketed by dashed lines.
}
\end{center}
\end{figure}

\clearpage

\begin{figure}
\begin{center}
\includegraphics[scale=0.8]{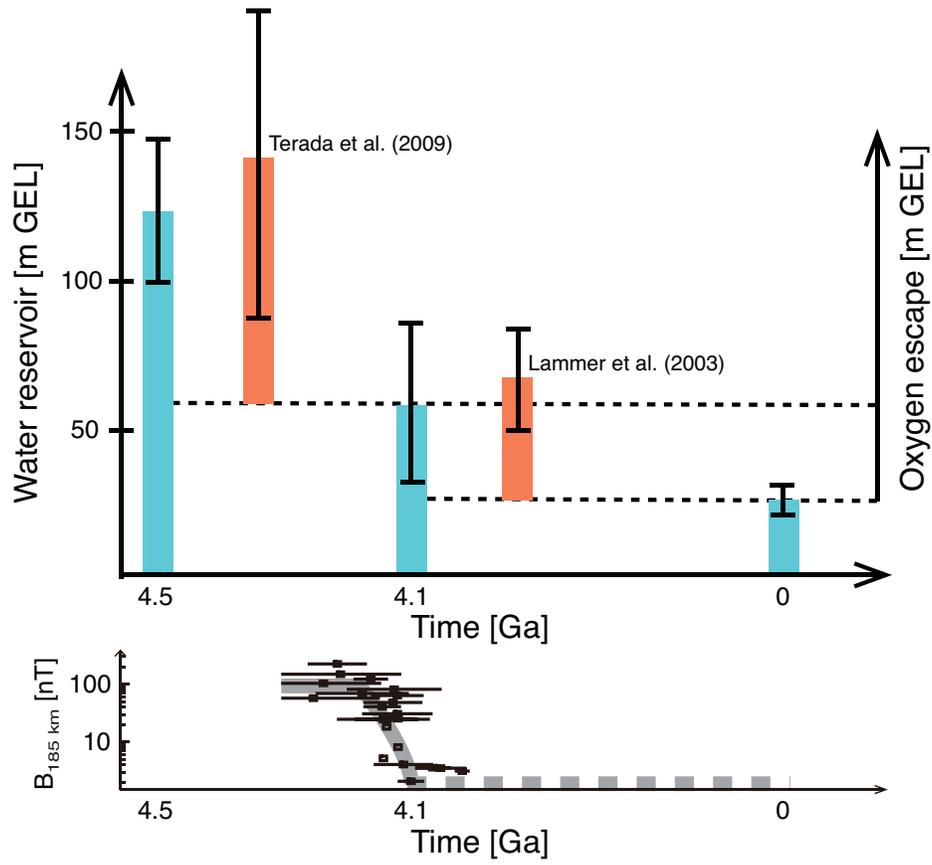}
\caption{
Upper diagram: evolution of martian water reservoir estimated from the minimum amount of the present water reservoir in PLD (blue).
The error bars are derived from the uncertainty of the amount of present water reservoir $R_{\rm present}$ ($20 - 30\ {\rm m}$) and the ${\rm \delta D}$ value ($1200 - 3000$ñ) for ALH 84001.
The comparisons with water loss estimated by oxygen escape calculation models are also shown (orange); \citet{Lammer+2003} ($14 - 34\ {\rm m}$ GEL) and \citet{Terada+2009} ($18 - 78\ {\rm m}$ GEL).
Lower diagram: magnetic field observed over large basins are plotted as function of N(300) crater age of \citet{Frey2008} (modified after \citet{Lillis+2008}). $B_{\rm 185 km}$ is magnetic field magnitude at 185 km altitude above the martian datum.
}
\end{center}
\end{figure}

\clearpage

\begin{figure}
\begin{center}
\includegraphics[scale=1.0]{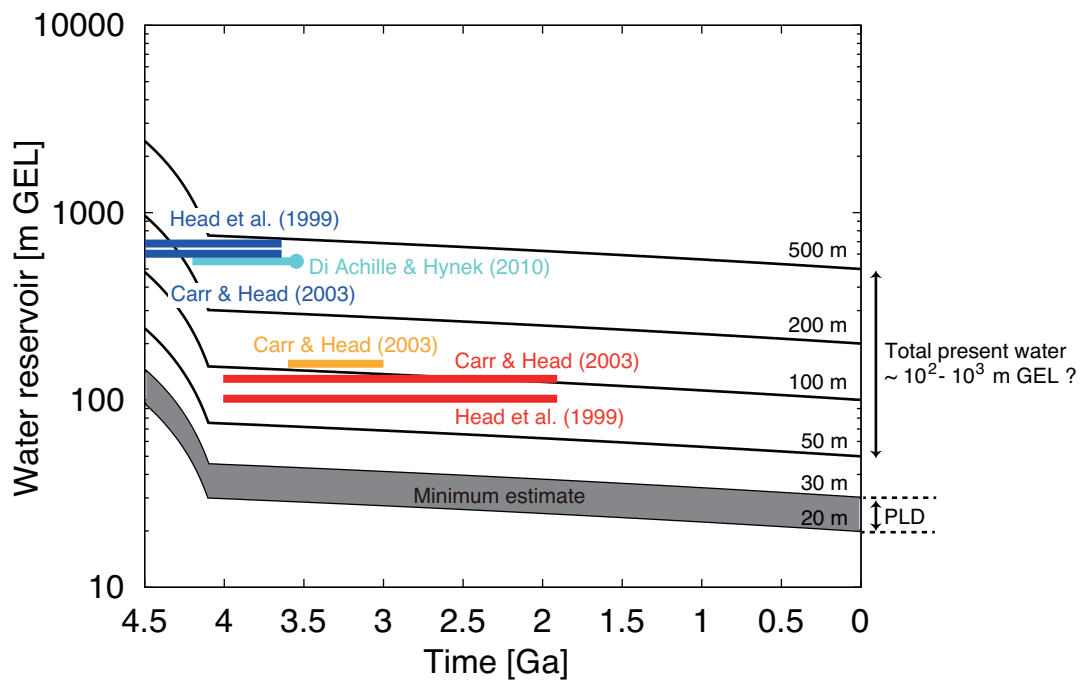}
\caption{
Evolution of water reservoirs for different amounts of present water reservoirs (black lines; $20,\ 30,\ 50,\ 100,\ 200,\ {\rm and}\ 500\ {\rm m\ GEL}$) and geological estimates of water amount: Contact-1 and Arabia shoreline (dark blue), Noachian ocean based on delta and valleys (light blue), Late-Hesperian ocean based on VBF (orange), and Contact-2 and Deuteronilus shoreline (red). 
The gray area indicates the evolution of surface water reservoir calculated based on the minimum present water reservoir ($20-30\ {\rm m}$ GEL) estimated from PLD.
We assume ${\rm \delta D} = 3000$ñ at $4.1\ {\rm Ga}$ in this figure.
}
\end{center}
\end{figure}

\clearpage

\begin{table}[htb]
\begin{center}
\caption{Estimated water loss during Stage-1 and -2. Estimates based on oxygen escape calculations are from 1: \citet{Terada+2009} ($18 - 78\ {\rm m}$ GEL in the original paper with different conversion) and 2: \citet{Lammer+2003} ($14 - 34\ {\rm m}$ GEL in the original paper with different conversion). See text for details.}
\begin{tabular}{lrr} \hline
Method & Stage-1 & Stage-2 \\ \hline
Based on PLD & $41-99\ {\rm m\ GEL}$ & $10-53\ {\rm m\ GEL}$ \\
Based on \citet{Carr+Head2003} & $53-280\ {\rm m\ GEL}$ & $120-520\ {\rm m\ GEL}$ \\
Oxygen escape & $31-133\ {\rm m\ GEL}$\footnotemark & $24-58\ {\rm m\ GEL}\footnotemark$ \\ \hline
\end{tabular}
\end{center}
\end{table}
\footnotetext[1]{\citet{Terada+2009}. Originally $18-78$ m GEL with a different conversion in which a $1$ m GEL ocean contains $8 \times 10^{42}$ water molecules ($= 2.4 \times 10^{17}$ kg).}
\footnotetext[2]{\citet{Lammer+2003}. Originally $14-34$ m GEL in with a different conversion in which a $1$ m GEL ocean contains $8 \times 10^{42}$ water molecules ($= 2.4 \times 10^{17}$ kg).}
\clearpage

\bibliographystyle{model2-names}
\bibliography{<your-bib-database>}



\end{document}